\documentclass{emulateapj}
\usepackage{apjfonts}

\newcommand{\gtsimeq}{\raisebox{-0.6ex}{$\,\stackrel 
        {\raisebox{-.2ex}{$\textstyle >$}}{\sim}\,$}}

\newcommand{\av}{$A_V$\,}
\newcommand{\ebv}{$E_{B-V}$\,}
\newcommand{\alfopt}{$\alpha_{\rm opt}$}

\shorttitle{No evidence of `gray' dust from composite quasar spectra}
\shortauthors{Willott}

\begin{document}


\title{No evidence of `gray' dust from composite quasar spectra}



\author{Chris J. Willott\altaffilmark{1}}

\altaffiltext{1}{Herzberg Institute of Astrophysics, National Research
Council, 5071 West Saanich Rd, Victoria, BC V9E 2E7, Canada;
chris.willott@nrc.ca} 

\begin{abstract}

Two recent studies based on composite reddened quasar spectra have
indicated the presence of `gray' dust in quasar environments. This
gray dust has a relatively flat extinction law in the UV, consistent
with the theoretical expectation of a lack of small dust grains close
to a quasar. In contrast, individual reddened quasars in the Sloan
Digital Sky Survey tend to have steep extinction laws in the UV,
similar to that in the SMC. We analyze the method used in determining
extinction laws from composite quasar spectra in order to resolve this
discrepancy. We show that quasars reddened by SMC-type dust that are
present in quasar samples have a negative correlation between
$E_{B-V}$ and redshift, due to selection effects. The fact that the
highest redshift quasars (which contribute to the UV part of a
composite spectrum) are less extincted leads to shallower extinction
in the UV. We construct a composite quasar spectrum from a simulated
sample of quasars reddened by SMC-type dust and show that the
extinction curve derived from the composite does not recover the
intrinsic extinction law. We conclude there is no evidence of gray
dust in quasar environments.

\end{abstract}


\keywords{dust, extinction  -- galaxies:$\>$active}

\section{Introduction}

Only $<0.01$\% of matter in galaxies is composed of dust --
coagulations of atoms and molecules with sizes ranging from a few
angstroms to a few microns (Seaquist et al. 2004). However, the
combined effects of extinction, scattering and thermal emission from
dust have a profound impact on our view of the universe. For example,
about half of the energy emitted as visible/UV radiation by stars in
distant galaxies has been reprocessed by dust into infrared emission.

Understanding the properties of this dust is vital. While dust in our
Galaxy and the Magellanic Clouds has been heavily studied (see,
e.g., the review by Draine 2003), there has been only limited work in
external galaxies, especially at high redshift. Of particular interest
is the possibility that `gray' dust is commonplace at high
redshifts. Gray dust is characterized by a flat extinction curve in
the UV and as such could be hard to identify from rest-frame UV data
alone (i.e. observed-frame optical for sources at high
redshift). Large surveys underway to measure cosmological parameters
from the Hubble diagram of Type Ia supernovae could be seriously
affected by the presence of unknown quantities of gray dust at high
redshift (Aguirre 1999).

One of the best ways to study dust extinction in distant galaxies is
via the effect that dust has on background quasars. Quasars provide a
bright UV source that can still be detected at high significance even
after significant absorption. However, a caveat is that if the dust
extinction occurs close to the active nucleus, then the unusual gas
density and radiation field could lead to atypical dust properties.

In the past, radio-loud quasar surveys have proved most fruitful for
searching for reddened quasars due to the fact that radio emission is
unaffected by dust (Webster et al. 1995; Baker \& Hunstead 1995; Gregg
et al. 2002; White et al. 2003). However the vast area and generous
color selection criteria employed by the Sloan Digital Sky Survey
(SDSS) mean that it has discovered hundreds of reddened quasars
(Richards et al. 2003). In an analysis of the continuum shapes of 9566
SDSS quasars, Hopkins et al. (2004) showed that the dust extinction is
best fit by dust similar to that in the SMC, i.e., extinction
increasing sharply in the UV. This work included 1886 SDSS quasars
with 2MASS matches which give the longer wavelength baseline necessary
to better distinguish between different extinction laws.

In stark contrast to the results from Hopkins et al. (2004), there
have been two recent studies that claim that the UV extinction curves
in quasars are `gray' or flatten in the UV (Czerny et al. 2004;
Gaskell et al. 2004). Both these works use the ratios of reddened and
unreddened composite quasars to derive quasar extinction laws. In this
Letter we analyze their methodology to understand the discrepancy
between these works and that of Hopkins et al.

\section{Using composite quasar spectra to derive extinction laws}

The existence of large digital quasar surveys has led to the now
commonplace technique of combining spectra of many quasars into a
composite spectrum (Francis et al. 1991; Baker \& Hunstead 1995; Zheng
et al. 1997; Brotherton et al. 2001; Vanden Berk et al. 2001). These
composites improve the signal-to-noise ratio and wavelength baseline and
also allow ``typical'' quasar properties to be determined. However,
the range in {\em rest-frame} wavelengths as a function of redshift
probed by optical spectroscopy means there may be very little overlap
between the quasars that formed the rest-frame optical and UV parts
of the spectrum. Given that luminosity and redshift are always
correlated in flux-limited samples, this means that there is also a
luminosity-dependence of the composite as a function of
wavelength. 

Czerny et al. (2004) used the five SDSS composites generated by
Richards et al. (2003) to determine the quasar extinction law.  These
composites were generated from quasars separated into five bins based
on the {\em relative} $g'-i'$ colors (the colors with respect to the
median as a function of redshift, see Richards et al. for more
details); composites 1 and 2 lie on the blue side of the $g'-i'$ distribution
and 3 and 4 lie on the red side. The quasars for composite 5 (the
``reddened composite'') were defined by a relative $g'-i'$ that was a
function of redshift to incorporate the redshift-dependent effect of
dust reddening on observed frame-colors.  Richards et al. considered
the asymmetric color distribution and the emission line properties of
composites 1-5 and concluded that the differences between 1-4 were due
to a combination of dust reddening and {\em intrinsic} spectral slopes
and that the intrinsic differences dominate. Czerny et al., however,
assume that the bluest quartile SDSS composite 1 represents an
unreddened quasar and composites 3-5 differ from 1 solely due to dust
reddening.  An extinction curve is generated by dividing composites
3-5 by composite 1 and averaging the result. For this reason alone, we
do not expect the Czerny et al. curve to truly represent dust
reddening.  As we will show, there is a further effect that casts
doubt on their results.

Gaskell et al. (2004) used composite quasars from Baker \& Hunstead
(1995) to determine a quasar extinction law. Baker \& Hunstead
generated composite quasar spectra for four groups of radio-loud
quasars from the Molonglo Quasar Survey (MQS). The groups were defined
by the ratio of the radio fluxes of the core and lobe, $\Re$. $\Re$ is
an orientation indicator and Baker \& Hunstead found that the
composite spectrum of low $\Re$ quasars is redder than the high $\Re$
composite, presumably due to dust reddening by a torus perpendicular
to the radio jet axis. Baker \& Hunstead also produced a composite
compact-steep-spectrum quasar and this also appeared highly reddened,
although in this case the explanation is more likely to be an
evolutionary effect. Gaskell et al. used a similar method to Czerny et
al. and compared the $\Re<0.1$ composite to the $\Re \geq 1$ composite
and the compact-steep-spectrum composite to the $0.1 \leq \Re < 1$
composite. The extinction curves from these two pairs agree very well.
Gaskell et al. also compared the $\Re \geq 1$ composite to the
UV-selected composite spectrum of the Large Bright Quasar Survey
(LBQS; Francis et al. 1991), which showed a similar extinction
curve. However, since the $\Re \geq 1$ composite is the least reddened
of all the radio-loud composites it is not clear that differences
between this and the LBQS composite can be attributed to reddening
rather than intrinsic spectral differences associated with the
different selection methods (radio-loud vs radio-quiet).
 
Both Czerny et al. and Gaskell et al.  found the result that the
extinction laws are flatter in the UV than the extinction laws of the
SMC, LMC or Milky Way.  Both extinction curves are plotted in
Fig.\,\ref{fig:gas} along with the curves for the SMC and Milky
Way. The curve of Gaskell et al. is flat at all wavelengths below
$0.4\,\mu$m. The Czerny et al. curve is similar to that of the SMC
down to $0.25\,\mu$m and below this it flattens to become much flatter
than that of either the SMC or the Milky Way at $<0.13\,\mu$m. These
flat UV extinction curves are particularly interesting since they
agree with theoretical predictions based on the grain size
distribution close to an AGN with the destruction of small grains
due to grain growth and sputtering in a high-density,
high-radiation environment (e.g. Maiolino et al. 2001a,b).

As discussed above, there are serious concerns about combining quasar
spectra that span a wide range of rest-frame wavelengths into a
composite. This is especially true when dealing with reddened quasars.
Richards et al. (2003) and White et al. (2003) showed that the
majority of the quasars with reddening \ebv$\gtsimeq 0.2$ in their
surveys (which are optical and optical+radio selected, respectively)
are found at redshifts $z<1$. Although similarly reddened quasars
do exist at higher redshifts they are less likely to be found in
quasar samples with an optical magnitude limit. We outline here
several reasons why, in practice, quasar surveys contain more highly
reddened quasars at lower redshifts:

\begin{figure}
\resizebox{0.48\textwidth}{!}{\includegraphics{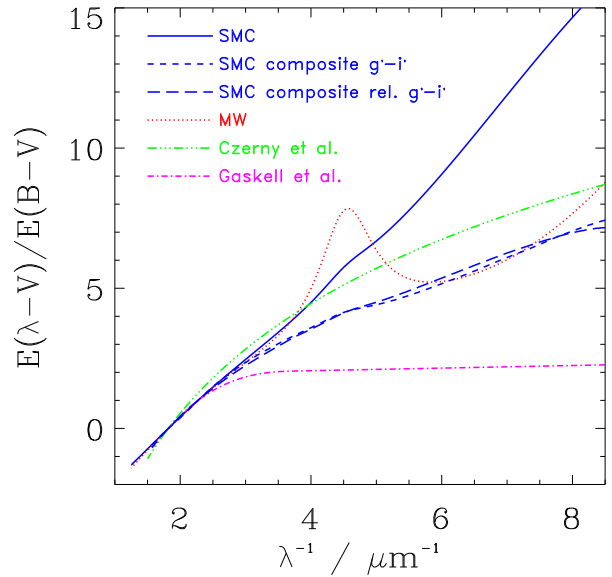}}
\caption{Wavelength dependence of extinction for various extinction
  laws: SMC (Pei 1992; {\it solid line}), Milky Way (MW; Pei 1992;
  {\it dotted line}), AGN extinction laws derived from composite
  quasar spectra (Czerny et al. 2004, {\it dash-triple-dotted line};
  Gaskell et al. 2004, {\it dash-dotted line}). The two dashed lines show the
  resulting extinction curves determined from a simulated sample of
  quasars reddened by the SMC law (Sec.\,3). It rapidly diverges from
  the SMC law in the UV due to the redshift-dependent extinction
  present in typical quasar samples.
\label{fig:gas}
}
\end{figure}

\begin{enumerate}

\item{At high redshift the selection is based on the rest-frame UV
  spectrum and $A_{UV}\,\gg$\,\av for SMC or Milky Way extinction
  laws. Therefore the \av necessary to exclude quasars from samples
  due to optical magnitude limits is a function of redshift.}

\item{There is mounting evidence of an increase in the fraction of
  obscured AGN at lower luminosities (e.g. Ueda et al. 2003; Simpson
  2005). Due to the optical magnitude limit, quasars at lower
  redshifts are typically less luminous. Therefore the ratio of the
  space densities of lightly reddened to unobscured quasars is
  greater at low redshifts than at high redshifts.}

\item{Low redshift, low luminosity quasars can have significant
  contamination of the red end of their spectra by light from the host
  galaxy. If this is not identified as host galaxy contamination it
  could lead to the quasar being classified as reddened. At high
  redshift and luminosities there is little contamination by the
  hosts. For example, the SDSS composite quasar of Vanden Berk et
  al. (2001) shows a spectral steepening longward of 0.5\,$\mu$m due to
  host galaxy contamination in low redshift quasars.}

\item{Red quasar selection is sometimes based on the observed frame
  color or spectral index. For an extinction law that induces
  spectral curvature (such as the SMC), the relationship between
  observed color or spectral index and \ebv is also a function of
  redshift, such that high-redshift quasars have an \ebv lower than
  that of low-redshift quasars with the same color. Richards et
  al. (2003) attempted to take account of this bias in their composite
  5 spectrum, however they did not for composites 3 and 4 which Czerny
  et al. also used to represent reddened quasars.}

\end{enumerate}
The consequence of all these points is that the quasars going into a
composite spectrum will have a negative correlation between \ebv and
redshift.

\begin{figure}
\resizebox{0.48\textwidth}{!}{\includegraphics{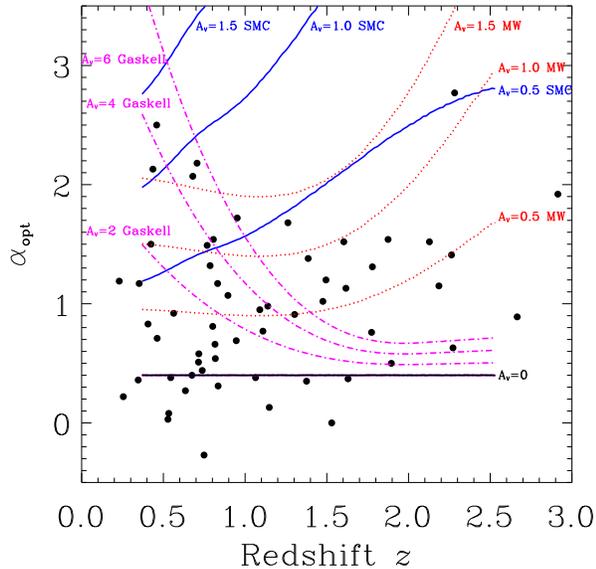}}
\caption{Optical spectral index, \alfopt, against redshift for the MQS
sample used to create the composites (Baker et al. 1999; {\it filled
circles}). Note that the reddest quasars in the observed-frame are
located preferentially at low redshifts. Also plotted are curves
showing how a power law in the observed-frame optical with
\alfopt\,$=0.4$ is affected by SMC ({\it solid line}) or Milky Way
({\it dotted line}) dust reddening as a function of redshift. This
shows that for a given \alfopt, the \av is greater for low redshift
quasars. Combining these two effects means that the \av (or \ebv) of
the quasars is a strong function of redshift in this sample. For the
Gaskell et al. extinction law ({\it dash-dotted line}), unfeasibly high values
of extinction are necessary to redden the high-redshift quasars.
\label{fig:mqs}
}
\end{figure}

To illustrate the significance of these effects in the light of the
current discussion, in Fig.\,\ref{fig:mqs} we plot optical spectral
index, \alfopt, against redshift for the MQS sample of Baker et
al. (1999) which was used to generate the composites in Baker \&
Hunstead (1995). It can be seen that in this sample, most of the
reddest quasars (\alfopt$>1.5$) are at lower redshifts. This is a
consequence of points 1--3 outlined above. However, there is not a
significant correlation between redshift and \alfopt\ in the full
sample because the bluest quasars are also predominantly at low
redshift.

The fact that there is no strong positive or negative correlation of
\alfopt\ with redshift does not mean however that there will be no
bias when making a composite from these quasars, because of point 4.
On Fig.\,\ref{fig:mqs} we also plot curves showing how the observed
spectral index changes with reddening as a function of redshift. For
simplicity we represent a quasar by a single power law with
\alfopt\,$=0.4$ and determine the observed \alfopt\ by the ratio of
fluxes in observed-frame wavelength bins $0.8-0.85\,\mu$m and
$0.35-0.4\,\mu$m (roughly corresponding to the extreme useful range of
optical spectroscopy). The power law is reddened by \av$=0.5,1$ and $1.5$
adopting the SMC and Milky Way (excluding the $0.22\,\mu$m feature not
usually seen in quasars) extinction laws (Pei 1992) and \av$=2,4$ and $6$
adopting the law of Gaskell et al. (2004). For the SMC and Milky Way
laws there is a strong correlation of \av with redshift in this
sample.  For the SMC dust, almost all the quasars with
$A_V>0.5$ are at $z<1$. Even for the Milky Way law which does not
induce such strong curvature, a correlation is seen at $z>2$.

This plot also illustrates a fundamental problem with the extinction
law of Gaskell et al. (2004). The extreme flatness in the UV means
that a huge \av is necessary to get any reddening at all of the UV
continuum. Many surveys, including the MQS and the SDSS, contain
reddened quasars at $z>2$ where only the UV is visible in optical
spectroscopy. According to the extinction law of Gaskell et al. the
quasars with \alfopt$\approx 1.5$ in the MQS must be reddened by
\av$\sim 20$ and therefore have intrinsic absolute magnitudes of
$M_B\sim -45$! What is actually found for high-redshift quasars in the
SDSS that have rest-frame optical data available is that the reddening
in the optical is much lower than in the UV, similar to SMC dust
(Hopkins et al. 2004).

\section{Simulating the biases of composite reddened quasars}

It is clear that combining the quasars in Fig.\,\ref{fig:mqs} into
composites could lead to serious problems for deriving a dust
extinction law. In particular, there is an inconsistency in the
extinction law of Gaskell et al. (derived from the MQS) and the
existence of red high-redshift quasars in the MQS. To understand this
quantitatively we now consider what would happen if one were to
generate a composite from a quasar sample selected on the basis of
their red observed optical colors (or equivalently, steep
\alfopt). This analysis only includes the bias due to point 4 of the
previous section. 

We consider a hypothetical sample of 211 quasars distributed smoothly
over the redshift range $0.4 \leq z \leq 2.5$. This redshift range is
typical of quasar samples such as the MQS and the SDSS study of
Richards et al. (2003). It is also the redshift range where the
observed-frame wavelengths of optical spectroscopy (assumed
$0.38-0.92\,\mu$m as in the SDSS) are covered by our composite
spectrum ($0.1-0.8\,\mu$m). We assume that the intrinsic spectrum of
each quasar is represented by a composite spectrum specifically
constructed from the SDSS to represent a typical unreddened quasar
(Willott et al. in prep.).

Each quasar is reddened by some amount of SMC type dust until its
observed $g'-i'$ color equals 0.7. We also repeated the analysis
using relative $g'-i'$ colors equal to 0.5. Note that the use of
relative colors takes out the effect of emission lines and is a
closer approximation to selecting quasars with a constant \alfopt. As
we will show, there is little difference in the results when using
$g'-i'$ or relative $g'-i'$. To get an idea of the amounts of
reddening required to produce these colors as a function of redshift,
\av$=0.8$ at $z=0.4$ and \av$=0.19$ at $z=2.5$ for relative
$g'-i'=0.5$. For reference to the MQS data in Fig.\,\ref{fig:mqs},
this corresponds to observed \alfopt\,$\approx\,1.5$. We repeated our
analysis with different values for the colors and found only minor
differences in the results.

The individual reddened quasars are then combined using the same
procedure as in Vanden Berk et al. (2001). Because we use the
geometric mean composite spectrum, the spectral shape, extinction law
and mean extinction should all be preserved in the absence of any
biases (Reichard et al. 2003).  Therefore we now divide the composite
reddened quasar spectrum by the intrinsic quasar spectrum to determine
the extinction curve in the same manner as Czerny et al. and Gaskell
et al.

The results are shown in Fig.\,\ref{fig:gas} where we plot the curves
generated from our two composites (using constant $g'-i'$ and constant
relative $g'-i'$ reddening). The first thing to notice is that our
derived composite extinction curves are nothing like the original SMC
curve which was used to redden the individual quasars.  The typically
lower reddening in the high redshift quasars that dominates the UV
part of the spectrum lead to a much flatter extinction curve than the
SMC. In particular, at $\lambda^{-1}> 6\,\mu{\rm m}^{-1}$, the curves
are much flatter than either the SMC or the Milky Way, which is a
characteristic of extinction curves generated from composite quasars.

Our simulated extinction curves are even flatter than that of Czerny
et al., but have a shape remarkably similar to theirs. The fact that
one of three reddened composites used by Czerny et al. was
selected taking account of this redshift dependent bias may account
for the fact that their curve is somewhat closer to the SMC curve than
our derived curves. Therefore we find that the difference between the
extinction laws of Czerny et al. and the SMC can be fully accounted
for by this bias.

The Gaskell et al. law is extremely flat in the UV, so if their curve
really does come from quasars extincted with SMC type dust then a
stronger bias than represented by an approximately constant \alfopt\,
must exist. We can only speculate on exactly why this is without full
rest-frame UV and optical spectroscopy of the MQS quasars that went into
the composites. We note that Baker \& Hunstead (1995) scaled their
quasar spectra before combining to a common flux-density at
0.3\,$\mu$m. If we repeat our analysis using this scaling, rather than
the method of ordering each quasar by its redshift and scaling to the
common wavelength region of the previous mean (Vanden Berk et
al. 2001), then we find somewhat flatter curves in the UV than shown
in Fig.\,\ref{fig:gas}. Also, we note that the number of quasars in
each MQS composite was small (in the range 13-18) and at the extreme
wavelength ranges of the composites even fewer quasars were
contributing. Therefore object-to-object differences could have had a
large impact. Gaskell et al. claimed that the fact that the broad
emission lines revealed the same extinction curve as the continuum was
independent evidence of the flat extinction curve. The broad
emission lines are subject to exactly the same bias as the continuum,
so it is not in fact independent.

\section{Discussion}

Our analysis shows that the gray UV extinction laws derived using
composite quasars are an erroneous artifact of the technique. This is
due to the fact that the quasars contributing to the composite in the
UV have typically lower reddening than those that contribute in the
optical.

There have been other studies that have claimed that gray dust may be
present in AGN environments. Maiolino et al. (2001a,b) presented a
number of arguments for a different grain size distribution, devoid of
small grains. However their arguments are based largely on
mid-infrared extinction (which is dominated by very large grains) and
the different observed gas-to-dust absorption ratios in Seyfert
nuclei. However, Willott et al. (2004) showed that this difference in
gas-to-dust absorption ratios is also present in X-ray selected
quasars whose rest-frame UV and optical reddening can be fit with SMC
or Milky Way extinction laws, not gray dust. 

We argue that evidence based on rest-frame UV and optical spectral
energy distributions for individual reddened quasars (as in Hopkins et
al. 2004) is much more reliable than extinction laws based on
composite spectra. We conclude that the SMC curve
remains a useful description of the reddening in quasar environments
and that there is little evidence supporting extinction curves that
flatten in the UV.

\acknowledgments Thanks to Jo Baker, Daniel Fryer, Matt Jarvis, Ross
McLure and the anonymous referee for assistance and useful discussions.


\begin{thebibliography}{}

\bibitem{001} Aguirre, A. N., 1999, ApJ, 525 583
\bibitem{002} Baker, J. C., \& Hunstead, R. W. 1995, ApJ, 452, L95
\bibitem{003} Baker, J. C., Hunstead, R. W., Kapahi, V. K., \& Subrahmanya,
  C. R. 1999, ApJS, 122, 29
\bibitem{004} Brotherton, M. S., Tran, H. D., Becker, R. H., Gregg, M. D.,
Laurent-Muehleisen, S. A., \& White, R. L. 2001, ApJ, 546, 775
\bibitem{005} Czerny, B., Li, J., Loska, Z., \& Szczerba, R. 2004, MNRAS,
  348L, 54
\bibitem{006} Draine, B. T. 2003, ARA\&A, 41, 241 
\bibitem{007} Francis, P. J., Hewett, P. C., Foltz, C. B., Chaffee,
  F. H., Weymann, R. J., \& Morris, S. L., 1991, ApJ, 373, 465
\bibitem{008} Gaskell, C. M., Goosmann, R. W., Antonucci, R. R. J.,
\&  Whysong, D. H. 2004, ApJ, 616, 147
\bibitem{009} Gregg, M. D., Lacy, M., White, R. L., Glikman, E., Helfand, D.,
  Becker, R. H., \& Brotherton, M. S. 2002, ApJ, 564, 133
\bibitem{010} Hopkins, P. F., et al. 2004, AJ, 128, 1112
\bibitem{011} Maiolino, R., Marconi, A., Salvati, M., Risaliti, G.,
Severgnini, P., Oliva, E., La Franca, F., \& Vanzi, L. 2001a, A\&A, 365, 28
\bibitem{012} Maiolino, R., Marconi, A., \& Oliva, E. 2001b, A\&A, 365, 37
\bibitem{013} Pei, Y. C. 1992, ApJ, 395, 130
\bibitem{014} Reichard, T. A., et al. 2003, AJ, 126, 2594
\bibitem{015} Richards, G. T., et al.  2003, AJ, 126, 1131
\bibitem{016} Seaquist, E., Yao, L., Dunne, L., \& Cameron, H. 2004,
  MNRAS, 349, 1428
\bibitem{017} Simpson, C. 2005, MNRAS, 360, 565
\bibitem{018} Ueda, Y., Akiyama, M., Ohta, K., \& Miyaji, T. 2003,
  ApJ, 598, 886
\bibitem{019} Vanden Berk, D. E., et al. 2001, AJ, 122, 549
\bibitem{020} Webster, R. L., Francis, P. J., Peterson, B. A., Drinkwater, M. J., \& Masci, F. J. 1995, Nature, 375, 469
\bibitem{021} White, R. L., Helfand, D. J., Becker, R. H., Gregg, M. D., Postman,
M., Lauer, T. R., \& Oegerle, W. 2003, AJ, 126, 706
\bibitem{022} Willott, C. J., et al. 2004, ApJ, 610, 140 
\bibitem{023} Zheng, W., Kriss, G. A., Telfer, R. C., Grimes, J. P.,
  \& Davidsen, A. F. 1997, ApJ, 475, 469
\end{thebibliography}
\end{document}